\documentclass[aps,twocolumn]{revtex4}

\usepackage{graphics}
\begin{document}

\bibliographystyle{apsrev}

\title{Gravitational Waves from Spinning Compact Binaries}
\author{Neil J. Cornish}
\affiliation{Department of Physics, Montana State University, Bozeman, MT 59717}
\author{Janna Levin}
\affiliation{DAMTP, Cambridge University, Wilberforce Road, Cambridge CB3 0WA, United Kingdom}

\begin{abstract}
Binary systems of rapidly spinning compact objects, such as black
holes or neutron stars, are prime targets for gravitational wave
astronomers. The dynamics of these systems can be very complicated due
to spin-orbit and spin-spin couplings. Contradictory results have been
presented as to the nature of the dynamics.  Here we confirm that the
dynamics - as described by the second post-Newtonian approximation to
general relativity - is chaotic, despite claims to the contrary.
When dissipation due to
higher order radiation
reaction terms are included, the chaos is dampened. However, the
inspiral-to-plunge transition that occurs toward the end of the
orbital evolution does retain an imprint of the chaotic behaviour.

\end{abstract}

\maketitle

Gravitational wave astronomy blurs the lines between theory and
observation by requiring accurate source modeling to facilitate
detection. While it is possible to detect gravitational waves without
precise waveform templates, matched filtering against a template bank
is the only way to extract detailed information about the sources. A
template based, matched filtering approach to gravitational wave data
analysis is impractical if the orbital dynamics is chaotic
\cite{{ms},{jj1},{njc2},{jj2}}. Systems
that exhibit sensitive dependence to initial conditions require
template banks that are exponentially larger than those of non-chaotic
systems.

Spinning compact binaries pose a challenge to template based detection
and parameter extraction techniques. The waveforms depend on a large
number of parameters, including the masses of the two bodies, their
spins, and the relative alignment of the spin and orbital angular
momentum - some 11 parameters in all. Even with a relatively coarse
sampling of parameter space, the resulting template bank can be very
large. The hope is that hierarchical schemes can be used that start
with a coarse sampling and proceed by successive refinement. However,
template based methods, hierarchical or otherwise, will not work if
the underlying dynamics is chaotic. The sensitivity to initial
conditions that characterizes chaotic systems ensures that waveforms
that are initially nearby (as measured by their cross correlation over
some time interval) will diverge exponentially with time \cite{njc2}.

A debate has arisen as to whether spinning compact binaries exhibit
chaotic behaviour.  The first indication of chaotic behaviour was found
using a test particle approximation \cite{ms}, but chaotic orbits were
only found for unphysically large values of the particle's spin. The
problem was also approached using the post-Newtonian approximation to
general relativity, and fractal methods were used to show that
binaries with realistic spins exhibited chaotic
behaviour~\cite{jj1}. Commentaries were written emphasizing that
radiation reaction would damp the chaos~\cite{njc1}, and that the
post-Newtonian approximation was being pushed beyond its domain of
validity~\cite{scott}, although unavoidably since no better approximation is available.
However, neither commentary disputed the
central result of Refs.~\cite{{jj1},{jj2}} - that the second post-Newtonian
(2PN) equations of motion admit chaotic behaviour. Then a paper was
published ``ruling out chaos in compact binary
systems''\cite{ras}. This study used the same 2PN equations of motion,
but a different method for establishing chaos - Lyapunov exponents
rather than fractals. The results reported in Refs.\cite{jj1}
and \cite{ras} sit in stark contrast. The trajectories that form the
fractal basin boundaries found in \cite{jj1} belong to a set of
unstable periodic orbits know as the strange repellor. These orbits
must have positive Lyapunov exponents. Trajectories near the
boundaries will also have positive Lyapunov exponents, as may orbits
that lie far from the boundaries.

In what follows, we refute the claims made in Ref.\cite{ras} by
showing that the 2PN equations of motion do admit orbits with positive
Lyapunov exponents. We then explore the significance of this result by
comparing three key timescales in the problem - the average orbital
period $T_o$, the Lyapunov time $T_\lambda$, and the decay time $T_d$.
If $T_\lambda$ is short compared to $T_d$, the chaotic dynamics seen
in the conservative 2PN dynamics will leave a strong imprint on the
2.5PN dissipative dynamics.

The post-Newtonian equations of motion are written as a series
expansion in $v^2/c^2$, where $v$ is the relative velocity and $c$ is
the speed of light:
\begin{eqnarray}
\mu \ddot{\bf r} &=& {\bf F}_N^{(0)}+{\bf F}_{PN}^{(1)}+{\bf
F}_{SO}^{(1.5)}+{\bf F}_{PN}^{(2)} +{\bf F}_{SS}^{(2)}+{\bf
F}_{QM}^{(2)}\nonumber \\ && +{\bf F}_{RR}^{(2.5)}+{\bf
F}_{SO}^{(2.5)}+{\bf F}_{PN}^{(3)}+{\bf F}_{SS}^{(3)}+\dots
\end{eqnarray}
Here $\mu = m_1 m_2/M$ is the reduced mass, $M=m_1 +m_2$ is the total
mass, and $\ddot{\bf r}$ is the relative acceleration of the two
bodies. The product $\mu \ddot{\bf r}$ is given in terms of a series
of forces, starting with the usual Newtonian force ${\bf F}_N^{(0)} =
m_1 m_2 {\bf r}/r^3$. The superscripts denote the order of the
post-Newtonian expansion and the subscripts denote the type of
force. The explicit form of the higher order terms can be found in
Refs.\cite{{kidder},{poisson},{too}}.  Qualitatively, the 1PN force ${\bf
F}_{PN}^{(1)}$ introduces perihelion precession.  The 2PN force ${\bf
F}_{PN}^{(2)}$ introduces isolated unstable orbits, along with an
inner most stable circular orbit (ISCO), and the possibility of
merger. The 1.5PN spin-orbit force ${\bf F}_{SO}^{(1.5)}$ leads to
precession of the orbital plane, as do the 2PN spin-spin ${\bf
F}_{SS}^{(2)}$ and spin-induced quadrupole-monopole ${\bf
F}_{QM}^{(2)}$ forces. The spin-spin force is attractive for spins
that are aligned and repulsive for spins that are anti-aligned. The
2.5PN order radiation reaction force, ${\bf F}_{RR}^{(2.5)}$, is the
first non-conservative term, and it causes the orbital energy $E$ and
angular momentum $\bf L$ to decay. Associated with the spin-orbit and
spin-spin forces are torques that act on the spin of each body,
causing the spins to precess - see Ref.\cite{kidder} for
details. While the expansion is known to 3PN order, we will only
consider terms up to 2.5PN order to facilitate comparison with the
results in Refs.\cite{jj1,ras}. For the same reason, we also neglect
the 2PN quadrupole-monopole and 2.5PN spin-orbit forces. In defense of
these approximations, we point out that the terms that we keep capture
the main {\em qualitative} features expected from full general
relativity. If anything, the higher order non-dissipative terms are
likely to increase the strength of the chaotic behaviour.

The dynamics takes place in a 12-dimensional phase space with
coordinates $\vec{X}=( {\bf x}, {\bf p}_x, {\bf S}_1, {\bf S}_2)$,
where ${\bf p}_x$ is the momentum conjugate to ${\bf x}$, and ${\bf
S}_i$ describes the spins of the two bodies. In the absence of
radiation reaction there are 6 conserved quantities, the energy $E$,
total angular momentum ${\bf J}={\bf L}+{\bf S}_1+{\bf S}_2$ and spin
magnitudes $\vert {\bf S}_i \vert$. 
Linearizing the equations of motion about a reference trajectory $\vec X(t)$
gives the evolution of the difference ${\delta \vec X}(t)$
\begin{equation}\label{dev}
\delta \dot{X}_i(t) = \left. \frac{\partial \dot{X}_i}{\partial
X_j}\right\vert_{\vec{Y}(t)} \! \delta X_j(t) \equiv K_{ij}(t)\delta
X_j(t) \, .
\end{equation}
The solution to this equation can be written:
\begin{equation}\label{dx}
\delta X_i(t) = L_{ij}(t) \delta X_j(0) \, .
\end{equation}
The evolution matrix $L_{ij}(t)$ is given in terms of the linear
stability matrix 
$K_{ij}$ by 
\begin{equation}\label{levol}
\dot L_{ij}= K_{il} L_{lj}\, ,
\end{equation}
with $L_{ij}(0) = \delta_{ij}$ (repeated indices imply summation).  The
Lyapunov exponents are defined in terms of the eigenvalues
$\Lambda_i(t)$ of the distortion matrix $\Lambda_{ij}=L_{il}L_{lj}$:
\begin{equation}
\lambda_i = \lim_{t \rightarrow \infty} \frac{1}{2t} \log \Lambda_i(t)
\, .
\end{equation}
The 2PN equations of motion are conservative and can be derived from a
Hamiltonian.  The expansion and vorticity of the flow vanishes for
Hamiltonian systems (in canonical coordinates), so that ${\rm
det}(\Lambda_{ij})=1$, $\Lambda_{ij}=\Lambda_{ji}$ and the Lyapunov
exponents come in $+/-$ pairs that measure the exponential shearing of
the flow. The principal Lyapunov exponent $\lambda_p = {\rm
max}(\lambda_i)$ can be calculated without directly isolating the
eigenvalues from
\begin{equation}\label{L}
\lambda_p = \lim_{t \rightarrow \infty} \frac{1}{2t} \log \left(
\frac{ \Lambda_{jj} (t)}{\Lambda_{jj} (0)} \right) \, .
\end{equation}
In the limit of very long times, the principal positive Lyapunov
exponent will dominate the trace in eqn.\ \ref{L}.

By contrast, the quantity calculated 
in Ref.\cite{ras} was
\begin{equation}\label{dt}
\lambda_d = \lim_{t \rightarrow \infty} \lim_{dX(0)\rightarrow 0}
\frac{1}{t} \log \left( \frac{ dX(t)}{dX(0)} \right)
\end{equation}
with $dX=((X_i(t)-Y_i(t))(X_i(t)-Y_i(t)))^{1/2}$
the Cartesian distance between the 12-component
vectors of a reference trajectory $\vec X(t)$ and a
nearby shadow trajectory $\vec Y(t)$.
It must be emphasized that this is
{\it not} a Lyapunov exponent. Eqn.\ \ref{dt} will automatically yield
zero when the limit $t\rightarrow \infty$ is taken. However, eqn.\ \ref{dt} can 
represent an {\it approximation} to the Lyapunov exponent if an additional
rescaling of the shadow trajectories is incorporated. The rescaling is
accomplished by determining when $dX(t_r) > R\, dX(0)$ for some threshold $R$, then
starting a new shadow trajectory $\vec Y'(t)$ with initial conditions
\begin{equation}
\vec Y'(t_r) = \vec X(t_r) + \left(\vec Y(t_r)-\vec X(t_r)\right)/R \, .
\end{equation}
The rescaling is repeated throughout the evolution to ensure that one is accurately
approximating the stability of the reference trajectory $X(t)$. The
problem with this method is that the choice of threshold can
significantly affect the value of $\lambda_d$. 
It is possible that the apparent absence of rescaling in 
Ref.\ \cite{ras} is the source of our disagreement.
A far more robust method is to evolve the perturbation 
$\delta \vec{X}(t)$ directly using eqn.\ \ref{dev}.
No rescaling is needed as eqn.\ \ref{dev}
defines the dynamic stability without approximation.

A second more subtle point to make regarding eqn.\ \ref{dt} is that
while $dX(t)$ is often referred to as the ``distance between nearby
trajectories in phase space'', this statement is misleading as
phase space does not admit a metric structure. Even if rescaled properly
so that $dX(t)\approx \delta X(t)$ from eqn.\ \ref{dev},
the distance $dX(t)$ only
measures the projection of the distortion matrix onto the initial
displacement vector: 
\begin{equation}\label{Cd}
d^2X(t)\approx d^2(t)=dX_i(0) \Lambda_{ij}(t)dX_j(0). 
\end{equation}
As a consequence of this additional approximation,
$\lambda_d$ provides only a lower bound for
$\lambda_p$. 

\begin{figure}[ht]
\vspace{140mm} \includegraphics{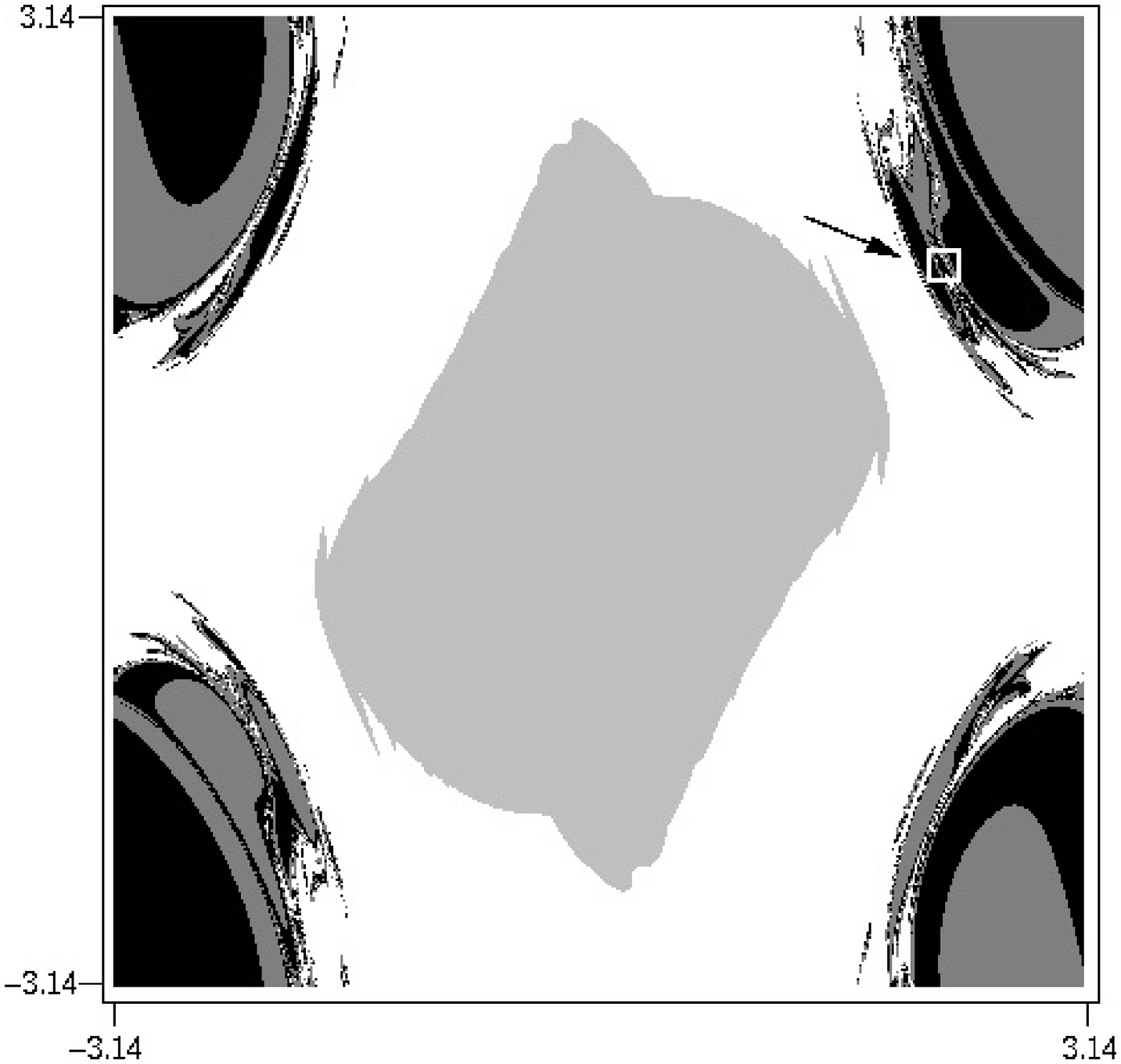} \includegraphics{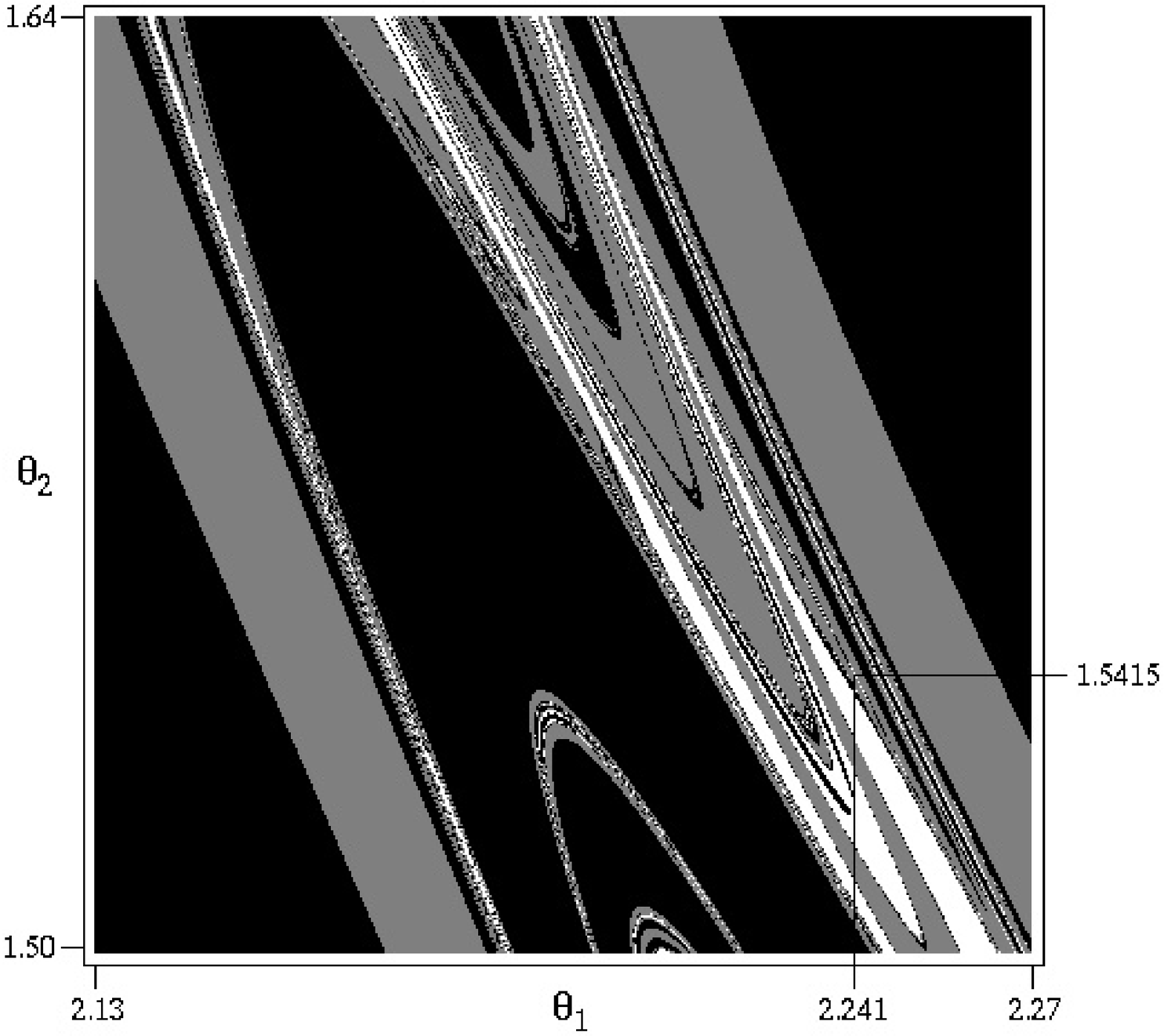}
\vspace{-10mm}
\caption{Fractal basin boundaries showing three possible outcomes for
the binary system as a function of the initial spin alignments.
\label{fbb}}
\end{figure} 

We use three methods to estimate the principle Lyapunov exponent:

Method (A) determines the full evolution matrix $L_{ij}$ from eqn.\ \ref{levol}
and uses eqn. \ref{L} to calculate $\lambda_p$. This is the most
numerically intensive method
as it involves integration of the 144 components of the
evolution matrix $L_{ij}$ 
as well as the 12 components of the trajectory itself.
The advantage of this method is that it yields an unambiguous
computation of the stability of an orbit with no approximations.

Method (B) uses shadow trajectories and
eqn.\ \ref{dt} with a careful rescaling of the shadow orbit.
This method involves the approximation of eqn.\ \ref{dt}, along with rescaling
and the projection described by eqn.\ \ref{Cd}.
 
Method (C) uses eqn.\ \ref{dev} to evolve
$\delta \vec{X}(t)$ along the reference trajectory $\vec X(t)$
so that a total of 24 equations are integrated and used to calculate
\begin{equation}\label{C}
\lambda_c = \lim_{t \rightarrow \infty} \lim_{\delta X(0)\rightarrow 0}
\frac{1}{t} \log \left( \frac{ \delta X(t)}{\delta X(0)} \right) .
\end{equation}
This method combines the accuracy of integrating the stability equations
with the approximation of projecting onto the distortion matrix as in eqn.\
\ref{Cd}.

In addition to these
methods we also studied the rate of phase decoherence in the waveforms
of the reference and shadow trajectories. According to
Ref.~\cite{njc2}, the phase difference $\vert \delta \Phi(t)\vert$
should grow as $e^{\lambda_p t}$. In summary, all four methods for
estimating $\lambda_p$ use some measure $D(t)$, where $D(t)$ is equal
to $\Lambda^{1/2}_{jj}(t)$, $dX(t)$, $d(t)$ or $\vert \delta \Phi(t) \vert$,
depending on the method. In each case, the quantity $D(t)$ will have
an initial power-law rise that is followed by exponential growth for
unstable orbits.

\begin{figure}[ht]
\vspace{60mm} \includegraphics{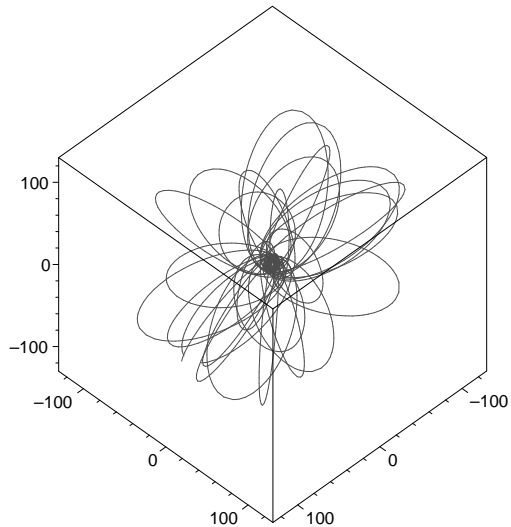}
\vspace{10mm}
\caption{Trajectory taken from the fractal basin boundary of Figure
1. The axes are scaled in units of the total mass.
\label{orb}}
\end{figure}

\begin{figure}[ht]
\vspace{60mm} 
\includegraphics{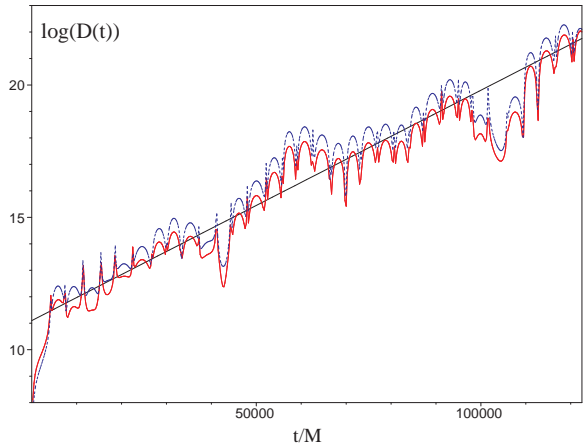}
\vspace{5mm}
\caption{Determining of the principle Lyapunov exponent for the
trajectory in Figure 2. The upper line uses method A, while the lower
two lines (which lie over one another) use methods B and C.
\label{le1}}
\end{figure}

To illustrate the connection between the fractal structures and
Lyapunov exponents, we begin by regenerating Figure 3 of
Ref.~\cite{jj1} in our Figure \ref{fbb}.  
The trajectories were started in the
$x-y$ plane with initial conditions $({\bf x},\dot{{\bf x}}) =(5.0 M
,0,0,0,0.45,0)$ and spin alignments $\theta_1$ and $\theta_2$
relative to the orbital angular momentum. The bodies have a $1:3$ mass
ratio and spins $S_i=0.6 m_i^2$. The trajectories were color coded
according to their outcomes: Black for merger from above the $x-y$
plane, dark grey for merger from below the $x-y$ plane, white for more than
50 orbits, and light grey for escape beyond $r=1000 M$. The lower
panel in Figure \ref{fbb} shows a detail of the fractal basin boundary, and
the location of a long-lived orbit that lies close to the basin
boundary. A portion of this trajectory is shown in Figure \ref{orb}.
The orbit has average period $T_o= 1687 M$, 
mean eccentricity $e=0.922$ and mean
semi-major axis $a=66.7 M$. Integrating the radiation reaction force
along the orbit gives a decay rate of $<\! \dot{E} \! >= -1.26 \times
10^{-6}$. In Figure \ref{le1} we plot $\log(D(t)/D(0))$ for this trajectory
using methods A, B and C described above. All three methods yield
$T_\lambda = 11500 M \sim 6.8 T_o$
for the Lyapunov
timescale, where we take an orbit to be a topological winding around the center
of mass. The Lyapunov timescale is less than seven orbital periods, indicating
that the motion is very chaotic.

\begin{figure}[ht]
\vspace{60mm} \includegraphics{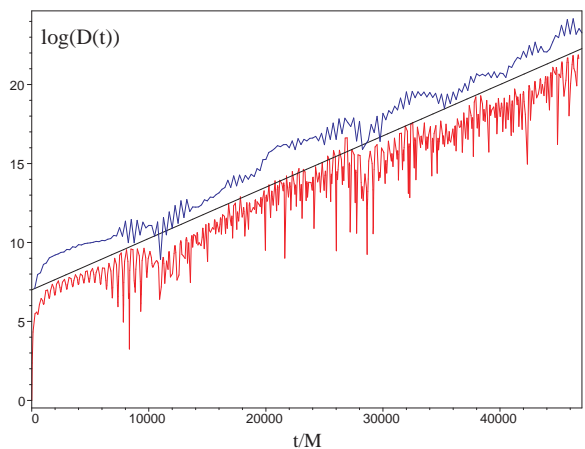}
\vspace{5mm}
\caption{Determination of the principle Lyapunov exponent for the less
eccentric orbit described in the text.  The upper line shows
$\log(d(t)/d(0))$ while the lower line shows $\log(\vert
\delta\Phi(t)\vert /\vert \delta\Phi(0)\vert)$.
\label{le2}}
\end{figure}

Similar results were found for many other orbits taken from Figure
\ref{fbb}. Most of the orbits near the boundaries tended to be highly
eccentric ($e>0.9$), by virtue of being on the boundary and so on the cusp
between merger and stability.
We did find some less eccentric orbits that
had positive Lyapunov exponents. For example, the trajectory with
initial conditions $({\bf x},\dot{{\bf x}}) =(5.5 M ,0,0,0,0.4,0)$,
$\theta_1 = \pi/2$, $\theta_2=\pi/6$, mass ratio $1:3$ and spins
$S_i=m_i^2$ is also highly chaotic. The orbit has 
average period $T_o=275 M$, 
mean eccentricity $e=0.59$ and mean semi-major axis
$a=13.7M$. Plots of $\log(D(t)/D(0))$ are shown in Figure \ref{le2}. In this
case we used method $C$ and the phase divergence method to estimate
$T_\lambda$. Both methods gave $T_\lambda = 3080 M = 11.2 T_o$, which
indicates that the orbit is highly chaotic.

We found large numbers of orbits, with a range of mass ratios, spin
parameters and spin alignments that had positive Lyapunov
exponents. The timescale for the
chaotic behaviour was often a small multiple of the orbital period. 

To further compare with Ref.\ \cite{ras}, we took up their case
of a binary with
mass ratio $1:1$ and spins $S_i=m_i^2$, $\theta_1=38^o$, $\theta_2=70^o$.
We found a positive Lyapunov exponent for
$({\bf x},\dot{{\bf x}}) =(5.0 M ,0,0,0,0.399,0)$. Therefore
at least some of
the equal mass binaries demonstrate chaotic orbits. As is common in 
chaotic systems, different trajectories came with different exponents, some
of which were zero. For example, the orbit with
initial conditions $({\bf x},\dot{{\bf x}}) =(5.0 M ,0,0,0,0.428,0)$
gave $\lambda_p=0$.
Herein lies an inherent weakness of the Lyapunov exponents themselves. 
They vary from orbit to orbit. A much more powerful survey of the phase space
scans for chaos using fractal basin boundaries as in Figure \ref{fbb}.

We used four methods to determine $\lambda_p$, along
with a battery of numerical tests, and our results
have proven robust. We therefore confirm  
the chaos discovered in Refs.\ \cite{{jj1},{jj2}}, contrary to the claims of
Ref.\ \cite{ras}.
It should also be emphasized that the fractal basin boundary method
used in Ref.\ \cite{jj1}
is an unambiguous declaration of chaos, and  
alone stands as proof of chaotic dynamics \cite{{njc},{mixm}}. 
Still, the Lyapunov
timescales can be useful for determining the impact of chaos on 
the gravitational wave detection.

\begin{figure}[ht]
\vspace{40mm} 
\includegraphics{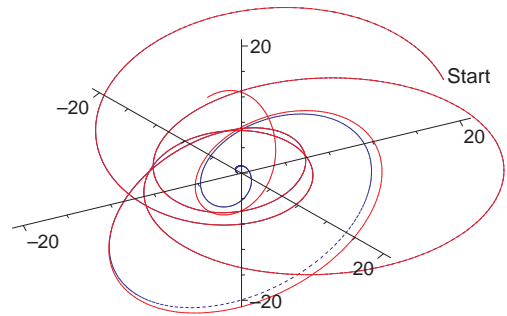}
%\special{psfile=odiss1.ps hscale=35 vscale=35 voffset=-70}
\vspace{5mm}
\caption{A detail of trajectory (solid line) showing the effect of
radiation reaction (dotted line).
\label{diss}}
\end{figure}

Now that we have confirmed that the 2PN dynamics is chaotic, we turn
to the question of how significant the effect is. To this end we went
to the next order in the post-Newtonian expansion and included the
radiation reaction force. The effect of the radiation reaction force
on the trajectory studied in Figure \ref{le2} is shown in Figure \ref{diss}. 
Starting at an arbitrary point along the orbit, we see
that the radiation reaction force drives the evolution from inspiral
to plunge in roughly 5 orbits. This is comparable to the Lyapunov timescale
of roughly 11 orbits. It tells us is that the chaotic
behaviour seen at 2PN order is marginal when radiation reaction is included. 
That is, chaos is damped by dissipation, but at least for some orbits the Lyapunov
timescale is comparable to the dissipation timescale. Both the instability of
the orbits and the degree of damping increase as merger is approached.

\begin{figure}[ht]
\vspace{130mm} \includegraphics{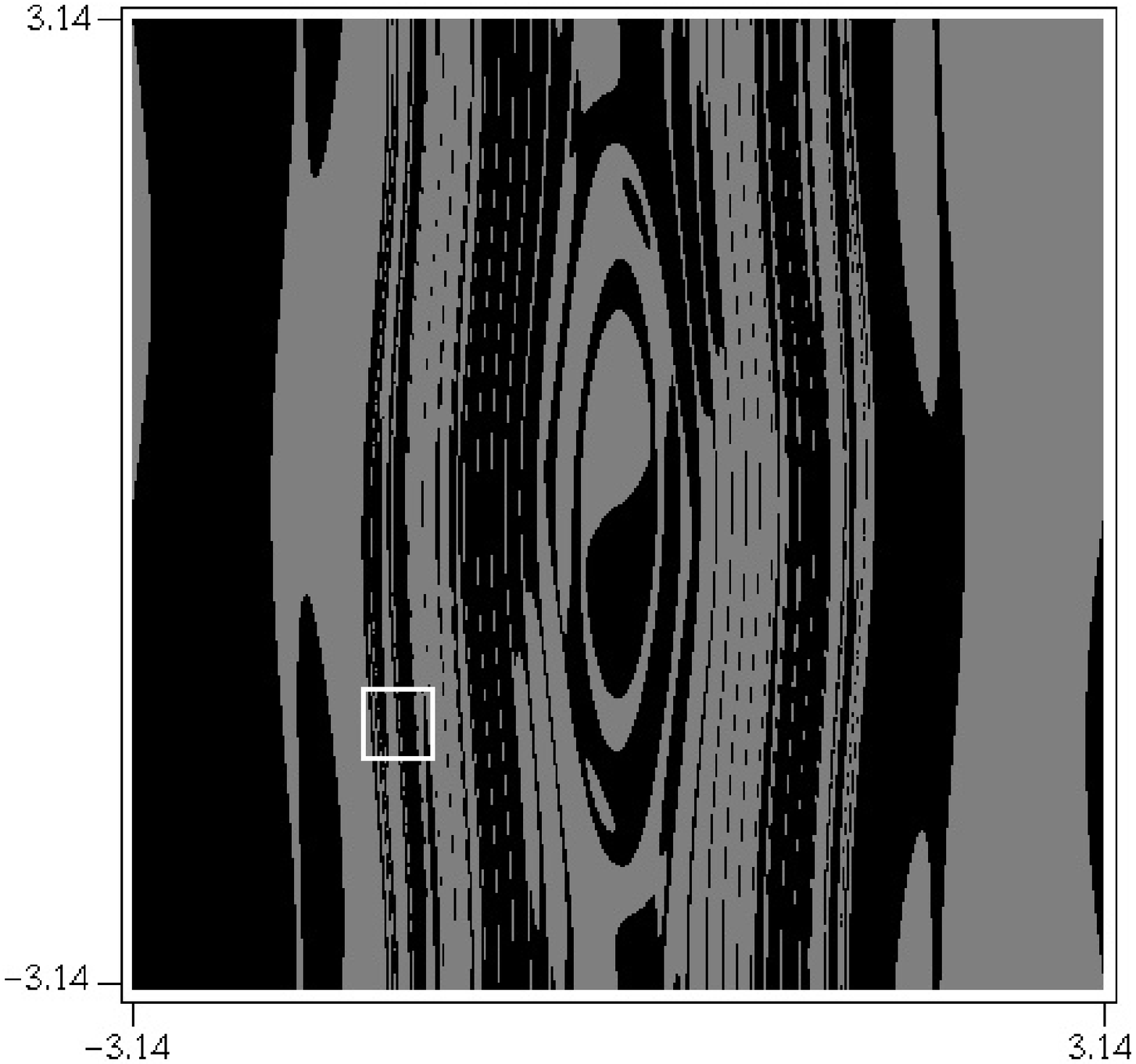}
\includegraphics{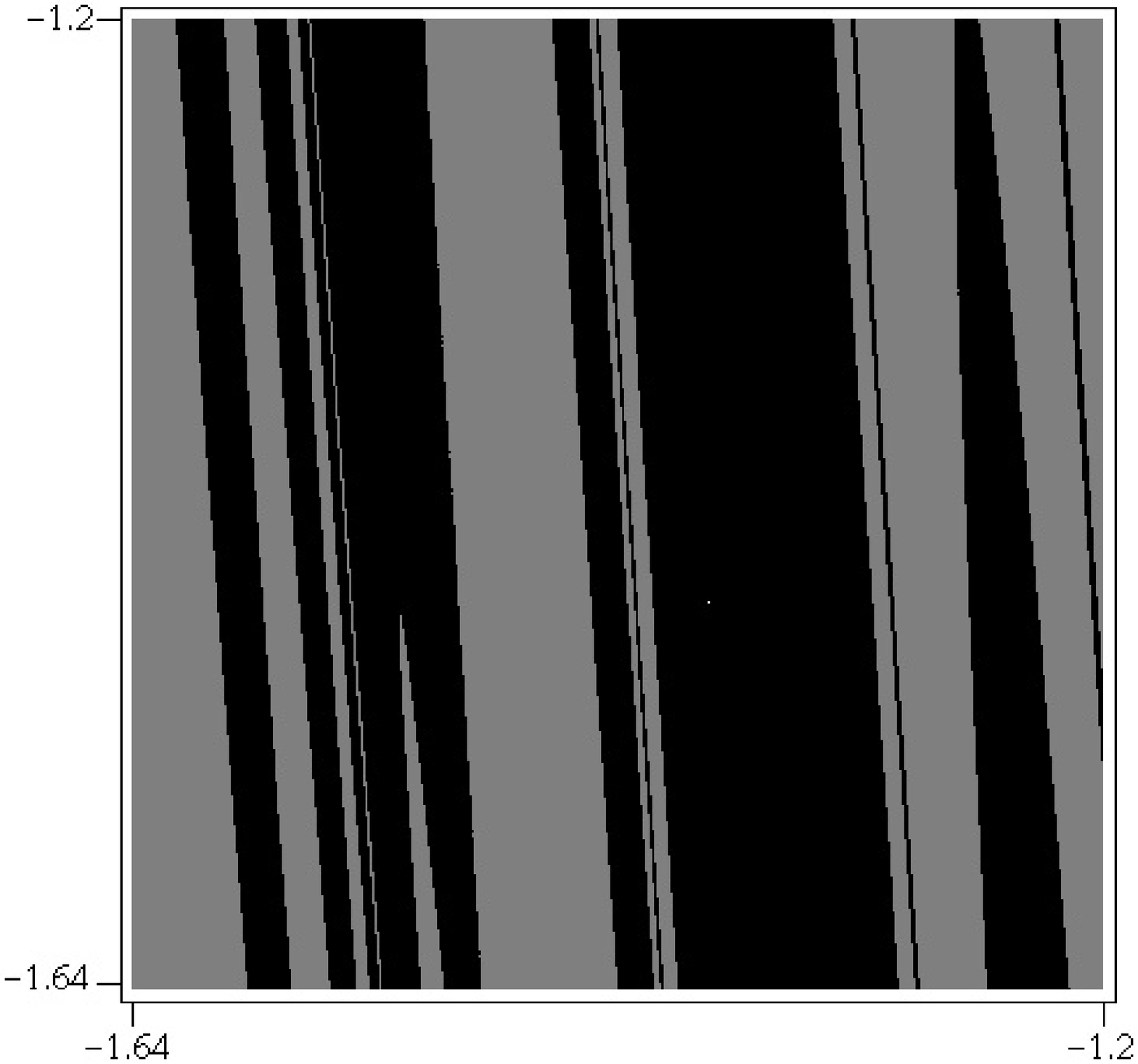}
%\vspace{8mm}
\caption{Basin boundaries with radiation reaction. The upper panel
shows a complicated intertwining of outcomes, however the detailed
view shows that the boundaries are not fractal.
\label{fbbd}}
\end{figure}

There is another way to show that the chaotic behaviour found in the
non-dissipative 2PN dynamics does leave an imprint on the dissipative
2.5PN dynamics. The effect is illustrated in Figure \ref{fbbd} where
trajectories with initial conditions $({\bf x},\dot{{\bf x}})=(30 M
,0,0,0,0.12,0)$, mass ratio $1:1$ and spins $S_i=0.6 m_i^2$ were
evolved for a range of spin-orbit alignments. The initial conditions
were color coded using the same scheme as before. Despite the damping,
the outcomes are intertwined in a complicated fashion. 
As pointed out in Ref.\ \cite{njc2}, with dissipation the system will
not show fully fractal boundaries. However, the imprint of 
the underlying chaos of the conservative system is recorded in the
amount of 
structure shown in the basin boundaries before the fractal cuts off and
is rendered smooth.
The detailed
view in the lower panel shows that the boundaries are
eventually smooth rather than fractal.

It is worth comparing our results to the interesting work of Ref.\ \cite{ms}.
The authors of Ref.\ \cite{ms} also found
a positive Lyapunov exponent for spinning test
particle motion around a Schwarzschild black hole. However, the light
companion required an unphysically large spin many many times maximal.
Here we find that the additional non-linearity from the gravitational 
interaction of the two bodies has introduced
chaotic dynamics for physically realistic spins below maximal. We emphasize
that the dynamics we study is only an approximation. We fully expect that
the additional corrections at higher order in the PN expansion will
augment the nonlinear behaviour and exacerbate the chaotic motion.
The very difficulty in solving the two-body problem in general relativity
hints that the two-body problem itself, perhaps even without
the addition of spins, is fully chaotic.

In conclusion, there is chaos in the 2PN equations of motion. The
chaos is damped by dissipation at 2.5PN order so that most orbits will only be  
mildly influenced by the complicated dynamics. What we draw from this
is that matched filtering may survive as a viable technique up until the
innermost orbits are reached. However, around the innermost orbits through to the plunge
we will need to rely on other techniques (as already noted in 
Ref.\ \cite{scott}). Importantly, the very idea of the innermost stable
circular orbit (ISCO) must be abandoned. It is telling that the most
unstable motion does appear to occur 
in the vicinity of the homoclinic orbits. (Homoclinic orbits
lie on the boundary between dynamical stability and instability
\cite{us}.  The ISCO is a specific
example of a homoclinic orbit.)
The underlying chaos of the
conservative dynamics means that unstable periodic orbits crowd this region 
of phase space. The fractal basin boundaries
are a reflection of this fractal set of unstable periodic orbits.
Consequently chaos can be significant for
the transition to plunge, as well as for the final orbits.

\section*{Acknowledgements}
NJC is supported in part by National Science Foundation Grant No. PHY-0099532.
JL is supported by a PPARC Advanced Fellowship.

\end{document}